\documentclass[twocolumn,showpacs,preprintnumbers,prl]{revtex4}
\usepackage{graphicx,bm,amsmath,amssymb}

\def\gz{\ifmmode{Z\hskip -4.8pt Z}
    \else{\hbox{$Z\hskip -4.8pt Z$}}\fi}

\newcommand{\be}{\begin{equation}}
\newcommand{\ee}{\end{equation}}
\newcommand{\bea}{\begin{eqnarray}}
\newcommand{\eea}{\end{eqnarray}}

\begin{document}

\title{Nonequilibrium transport through a singlet-triplet Anderson impurity}
\author{P.~Roura~Bas}
\affiliation{Centro At\'{o}mico Constituyentes, Comisi\'{o}n Nacional de Energ\'{\i}a 
At\'{o}mica, Buenos Aires, Argentina}
\author{A.~A.~Aligia}
\affiliation{Centro At\'{o}mico Bariloche and Instituto Balseiro, Comisi\'{o}n Nacional
de Energ\'{\i}a At\'{o}mica, 8400 Bariloche, Argentina}
\date{\today }
\date{\today}

\begin{abstract}
We study the Anderson model in which a configuration with a doublet is
hybridized with another with a singlet and a triplet. We calculate the
conductance through the system as a function of temperature and bias
voltage, near the quantum critical line for which the system is exactly
solvable. The results explain recent transport measurements in a
single-molecule quantum dot.
\end{abstract}

\maketitle

\email{roura@tandar.cnea.gov.ar}

\section{Introduction}

 When the wave function is forced to evolve continuously between two competing
ground states, a quantum phase transition (QPT) takes place between these two states 
At the transition, the length scale of quantum fluctuations becomes infinite,
and exotic states of condensed matter are expected \cite{sach}. 
Recently, several remarkable features of a (QPT) have been observed in
C$_{60}$ quantum dots (QD's) with even occupancy inserted in a nanoscale constriction \cite{roch}.
In the last years there has been a great interest in systems of quantum dots (QD's) 
because of the possible technological applications, and also because they
constitute ideal systems with a single magnetic impurity in which several
parameters can be tuned. 

When the QD has an odd number of electrons and the
Coulomb repulsion $U$ is large enough, the conductance at zero bias is
increased below a characteristic Kondo temperature $T_{K}$ as a consequence
of the Kondo effect. This is a usual feature of single-electron transistors
built with semiconductor QD's or single molecules \cite{gold,liang} 
and is well understood in terms of
the simplest ordinary Anderson model (OAM) in which a configuration with a
doublet is hybridized with a singlet. 
In a dot with an even number of electrons, there are two competing
states for the ground state configuration: a singlet in which two particles
occupy the lowest level (say $|00\rangle =s_{\uparrow }^{\dagger
}s_{\downarrow }^{\dagger }|0\rangle $) and a triplet in which one electron
is promoted to the next level and coupled ferromagnetically ($|11\rangle =p_{\uparrow }^{\dagger
}s_{\uparrow }^{\dagger }|0\rangle $ and its SU(2) partners) due to the
strong Hund coupling \cite{taru}. 
The
simplest Anderson model which describes the system mixes these four states
with a doublet ($|\sigma \rangle =s_{\sigma }^{\dagger }|0\rangle $)
by promoting a particle (electron or hole)
to one of the leads \cite{hof,paa}. This is the
singlet-triplet Anderson model (STAM) which had been used to describe valence fluctuating Tm
impurities in a cubic environment \cite{allub}. 

This model has a quantum phase
transition (QPT) from a singlet to a doublet ground state as the energy of
the triplet is decreased \cite{hof,allub}. When the triplet is well below
the other states, there is a partial screening of the spin 1 that explains
the zero bias Kondo peak observed experimentally in this situation \cite{sasa,schmid}. 
On the other (singlet) side of the transition there is a dip in the
conductance \cite{paa,hof} that has also been observed experimentally 
\cite{roch,paa}.

In this paper, we show that the experimental observations in C$_{60}$ QD's
can be understood in terms of this model and its QPT. 
The differential conductance $dI/dV$ as a function of temperature $T$ and bias voltage $V$
has been measured at both sides of the
transition \cite{roch}. On the singlet side of the transition, a dip in the
conductance at $V=0$ is observed in agreement with theoretical
expectations \cite{hof,paa} as well as non-equilibrium measurements 
performed in carbon
nanotubes \cite{paa}. On the other side of the transition, $dI/dV$ as a
function of $V$ shows a structure with three peaks that has not
been quantitatively explained yet. As the temperature $T$ is decreased, the zero
bias conductance $G(T)$ first increases, then shows a shoulder or a plateau
and then increases again. As stressed by the authors, this behavior is still not
understood \cite{roch}. The last two figures of our paper are the theoretical counterpart 
of these experimental results. The main qualitatively features are reproduced.
A more quantitative agreement would require a fine tuning of the parameters which 
is beyond the scope of this work. 
We provide an interpretation for the observed behavior.

\section{The model} 

The STAM assumes infinite $U$ and contains two neighboring charge
configurations in the QD: $d^{n}$ and $d^{n+1}$ with $n$ and $n+1$ particles.
Without loss of generality we can assume $n$ to be even,
performing an electron-hole transformation if necessary. Then, the $d^{n}$
configuration contains a singlet $|SM\rangle =|00\rangle $, 
where $S$ is the spin and $M$ its projection, and a triplet 
$|1M\rangle $, ($M=-1$, 0 or 1), while the $d^{n+1}$ configuration consists of a
doublet denoted by its spin 1/2 projection $|\sigma \rangle $. We introduce
the following creation operators for a particle in the dot \cite{note}

\begin{eqnarray}
d_{s\sigma }^{\dagger } &=&|\sigma \rangle \langle 00|,  \notag \\
d_{t\uparrow }^{\dagger } &=&-(|\uparrow \rangle \langle 10|+\sqrt{2}%
|\downarrow \rangle \langle 1-1|)/\sqrt{3},  \notag \\
d_{t\downarrow }^{\dagger } &=&(|\downarrow \rangle \langle 10|+\sqrt{2}%
|\uparrow \rangle \langle 11|)/\sqrt{3}.  \label{ope}
\end{eqnarray}%
The operators $d_{s\sigma }^{\dagger }$ and $d_{t\sigma }^{\dagger }$ 
hybridize via matrix elements $V^s_{\nu k}$ and $V^t_{\nu k}$ with the
conduction states $c_{\nu k \sigma }$ of two conducting leads $\nu=L$ (left) 
or $R$ (right) that transport the current
through the QD, leading to the Hamiltonian

\begin{eqnarray}
H &=&E_{s}|00\rangle \langle 00|+E_{t}\sum_{M}|1M\rangle \langle
1M|+E_{d}\sum_{\sigma}|\sigma \rangle \langle \sigma |  \notag \\
&&+\sum_{\nu k\sigma }\left[ (V^s_{\nu k}d_{s\sigma }^{\dagger }+V^t_{\nu k}d_{t\sigma
}^{\dagger })c_{\nu k\sigma }+\text{H.c.}\right]   \notag \\
&&+\sum_{\nu k\sigma }\epsilon _{\nu k}c_{\nu k\sigma }^{\dagger }c_{\nu k\sigma }.
\label{ham}
\end{eqnarray}

We assume $V^s_{L k}V^t_{R k}=V^s_{L k}V^t_{R k}$, so that only one conduction
channel $\sim \sum_{\nu k\sigma } V^{\eta}_{\nu k}c_{\nu k\sigma }$ 
($\eta=s$ or $t$) hybridizes with the dot states. In general, 
also the orthogonal linear combination of $c_{\nu k\sigma }$  
plays a role and ``screens''
the remaining doublet ground state when the localized triplet is well below the 
singlet, leading to a singlet ground state \cite{pus,pos}. 
However, the energy scale involved in this second screening $T^*$ (which depends 
exponentially on a small coupling constant \cite{pos}) might 
be very small. As discussed in Ref. \onlinecite{pos}, this is likely 
the case of previous experiments \cite{sasa,schmid}, as well as those in C$_{60}$ QD's \cite{roch}:
the theory in the general case \cite{pus,pos} predicts that the zero-bias conductance 
$G(T)$ should decrease at very low temperatures
and $dI/dV$ should also decrease for
the smallest applied bias voltages $V$ in contrast to the observations. 
This indicates that $T^*$ is
smaller than the smallest temperature in the experiments.

The STAM with only one conduction channel, 
also describes the mixing between the low lying states of the 
4f$^{12}$ and 4f$^{13}$ configurations in a cubic crystal field \cite{allub}.
For $E_{t}\rightarrow +\infty $, the model reduces to the OAM. For 
$E_{s}\rightarrow +\infty $, the model describes valence fluctuations between
two magnetic configurations \cite{bethe}. In both limits, 
for constant density of conduction states and hybridizations, 
the model is exactly
solvable (by the Bethe ansatz) and the ground state is a singlet (doublet)
in the first (second) case \cite{bethe}. Thus, the model has a QPT as a
function of $E_{s}-E_{t}$. The position of the transition 
depends on the other parameters of the model, leading to a quantum critical surface 
that can be determined calculating the magnetic
susceptibility at $T \rightarrow 0$ using numerical renormalization group (NRG) 
\cite{allub}. 
However, if $|V_{t}|^{2}=3|V_{s}|^{2}$, the transition takes place exactly
at $E_{s}-E_{t}=0$, independently of the value of $E_{d}$. In addition,
along this line, the model can be mapped into an OAM plus a free spin 1/2 
\cite{allub}. We will use these results to control the distance to the QCP.

\section{Approximations and equation for the current} 

As discussed for example in Ref. \onlinecite{none}, 
the calculation of non-equilibrium properties of a strongly correlated
system is a particular challenge for theory. A recent extension of the 
numerical renormalization group
(NRG) for the non-equilibrium case seems promising, but is not fully
developed yet \cite{and}. For the OAM, the non-crossing
approximation (NCA) \cite{win} and renormalized perturbation theory 
in $U$ \cite{hbo,rpt} have been useful. However, the latter 
is very difficult to extend to the STAM. The so called poor man's scaling works well when
either $eV$ or the magnetic field energy is larger than $kT_{K}$ \cite{rosch}
and has been successfully extended for a model similar to the STAM on the
singlet side of the QPT \cite{paa}. However, this method ceases to be valid
near the quantum critical surface for small $V$. 
Instead, the NCA can be extended to more general Anderson models,
like tose appropriate for Ce compounds \cite{ro1}, Co impurities on Ag and Cu \cite{ro2}
or systems of two quantum dots out of equilibrium \cite{agu}.

In this work, we extend the NCA to the STAM out of equilibrium. 
We introduce
auxiliary bosons, one for the singlet state and   three for the triplets,
and auxiliary fermions for the doublet, in analogy to the SU(N)$\times $%
SU(M) generalization of the Anderson model \cite{cox}. The spectral
densities of the operators $d_{\eta \sigma }^{\dagger }$  defined by 
Eqs. (\ref{ope}) for given spin, $\rho_{d}^{s}(\omega )$ and $\rho_{d}^{t}(\omega )$, 
are determined by
convolutions from those of the auxiliary particles. 
The current is given by \cite{meir}

\begin{equation}
I=\frac{A\pi e}{h}\int d\omega \lbrack \Gamma ^{s}\rho _{d}^{s}(\omega
)+\Gamma ^{t}\rho _{d}^{t}(\omega )][f_{L}(\omega )-f_{R}(\omega )],
\label{i}
\end{equation}
where $\Gamma ^{\eta }=\Gamma _{R}^{\eta }+\Gamma _{L}^{\eta }$ 
with $\Gamma _{\nu }^{\eta }=2\pi \sum_{k}
|V^{\eta}_{\nu k}|^{2}\delta (\omega -\epsilon _{k})$ assumed independent of
$\omega $ within a bandwidth $D$ and zero elsewhere, 
the asymmetry parameter $A=4\Gamma _{R}^{\eta }\Gamma
_{L}^{\eta }/(\Gamma _{R}^{\eta }+\Gamma _{L}^{\eta })^{2}$ 
(independent of $\eta $), and $f_{\nu }(\omega )$ is the Fermi function with
the chemical potential $\mu _{\nu }$ of the corresponding lead.

\section{Results}
 
For the numerical evaluations, we consider the case in which the ground state 
configuration has an even number of particles ($E_s,E_t <E_d$). 
We choose $E_{d}=(\mu _{L}+\mu _{R})=0$ (without loss of generality) and take 
a band width $D=10 \Gamma$,
$\Gamma ^{s}=1/2\Gamma $, and $\Gamma^{t}=3/2\Gamma $, where $\Gamma $ will be our unit of energy. 
The choice $\Gamma ^{t}/\Gamma ^{s}=3$ 
allows us to have an accurate control of the distance to the QPT, 
while the main features of the
results depend on this distance and not on the specific choice
of parameters.
For  $\Gamma ^{t}/\Gamma ^{s}=3$, if in addition $E_{s}=E_{t}$, the STAM
can be mapped into an OAM plus a free spin 1/2 (Ref. \cite{allub}). The OAM has
total coupling $\Gamma ^{\text{OAM}}=\Gamma $ and inverted charge transfer energy and
chemical potentials:  
$E_{d}^{\text{OAM}}-E_{s}^{\text{OAM}}=E_{s}-E_{d}$, $\mu _{\nu }^{\text{OAM}}=-\mu _{\nu }$. 
Using this mapping
it can be shown that the densities of both models are related by $\rho
_{d}^{s}(\omega )=\rho _{d}^{t}(\omega )=\rho _{d}^{\text{OAM}}(-\omega )/2$
and the absolute value of the current is the same. From the structure of the
corresponding NCA equations for both models, we realize that the NCA
satisfies these equalities. This has also been verified numerically.
Furthermore, for any $E_{d}$ the QPT takes place at $E_{s}=E_{t}$ when 
$\Gamma ^{t}/\Gamma ^{s}=3$ \cite{allub}. This defines a quantum critical line 
%(QCL) 
and  then $E_{s}-E_{t}$ controls the distance to this line. 

\begin{figure}[tbp]
%[rbp]
\includegraphics[width=7cm]{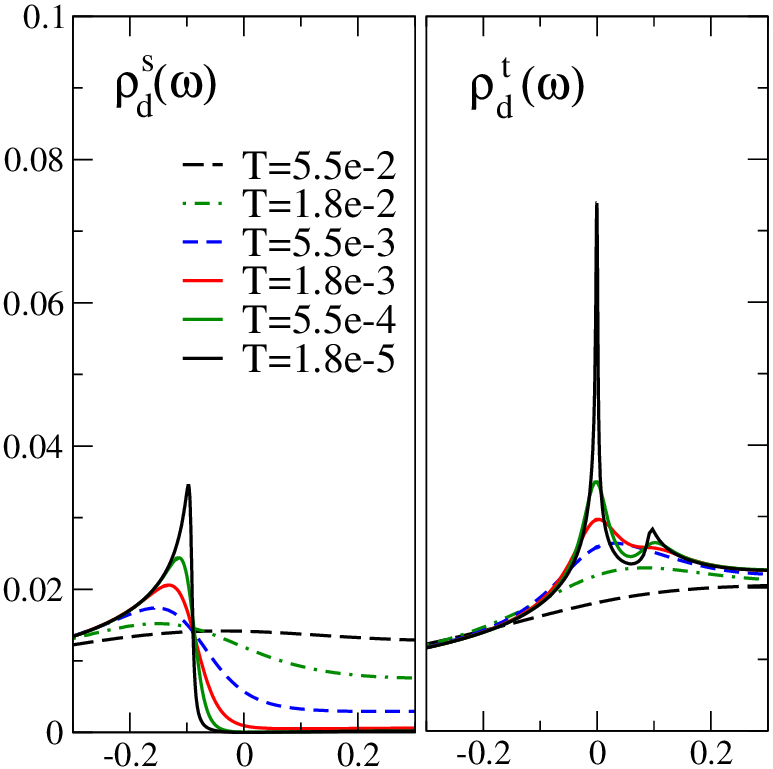}
\caption{(Color online) singlet (left) and triplet (right) contributions to the
dot spectral density for $E_t=-3, E_s=-2.9$ and different temperatures. 
The flatter curve corresponds to the highest temperature. 
$\Gamma=1$ is the unit of energy.}
\label{dens}
\end{figure}

As explained above, for $V=0$ ($\mu _{L}=\mu _{R}=0$),  when $E_{s}=E_{t}$, 
the singlet and triplet parts of the spectral 
density of the dot $2\rho _{d}^{s}(\omega )$ and $2\rho_{d}^{t}(\omega )$ 
coincide with the mirror image of the localized 
spectral density already reported
for the Anderson model (Fig. 5 of Ref. \cite{win} for $E_{t}=-2\Gamma $) and
exhibit the usual Kondo resonance at the Fermi level. 
The half width of this peak allows to define a Kondo temperature $T_{K}$.  
How do the spectral densities evolve as one
moves away from the quantum critical surface?  Decreasing $E_{s}$ 
(on the singlet side of the QPT) 
$\rho_{d}^{s}(\omega )$ displaces to positive frequencies, while 
$\rho_{d}^{t}(\omega )$ decreases and displaces its weight to negative 
frequencies. As a consequence, a gap opens in the sum 
$\Gamma ^{s}\rho_{d}^{s}(\omega )+\Gamma ^{t}\rho _{d}^{t}(\omega )$ 
entering Eq. (\ref{i}) at low temperatures. This situation has already been studied previously
and interpreted as the low temperature part of a
two-stage Kondo effect  \cite{hof}. 

The spectral densities for $E_{s}>E_{t}$
are shown in Fig. \ref{dens}. In contrast to the previous case, 
$\rho_{d}^{t}(\omega )$ remains peaked at the Fermi energy at low temperatures.
This is a consequence of the partial Kondo effect, by which the spin 1 at
the dot forms a ground state doublet with the conduction electrons of both
leads \cite{bethe}. The singlet part of the density $\rho _{d}^{s}(\omega )$ displaces to 
negative frequencies in this case. Therefore a pseudogap appears in the sum 
$\Gamma ^{s}\rho_{d}^{s}(\omega )+\Gamma ^{t}\rho _{d}^{t}(\omega )$, but at finite
frequencies, in contrast to the gap at the Fermi level that develops at the singlet side of 
the QPT. Note that at high temperatures both densities are quite similar and
the differentiation between $\rho _{d}^{s}(\omega )$ 
and $\rho _{d}^{t}(\omega )$ develops at a characteristic temperature of the order
of a fraction of $E_{s}-E_{t}$.

The equilibrium ($V=0$) conductance as a function of temperature $G(T)$ on the
singlet side of the QPT, near the quantum critical line (not shown) 
shows a maximum at a finite temperature and agrees with previous results using NRG \cite{hof}
and with experiment \cite{roch}. In particular, the increase and decrease of 
$G(T)$ from its maximum value are logarithmic to a good degree of accuracy.

\begin{figure}[tbp]
\vspace{1cm}
%[rbp]
\includegraphics[width=6cm]{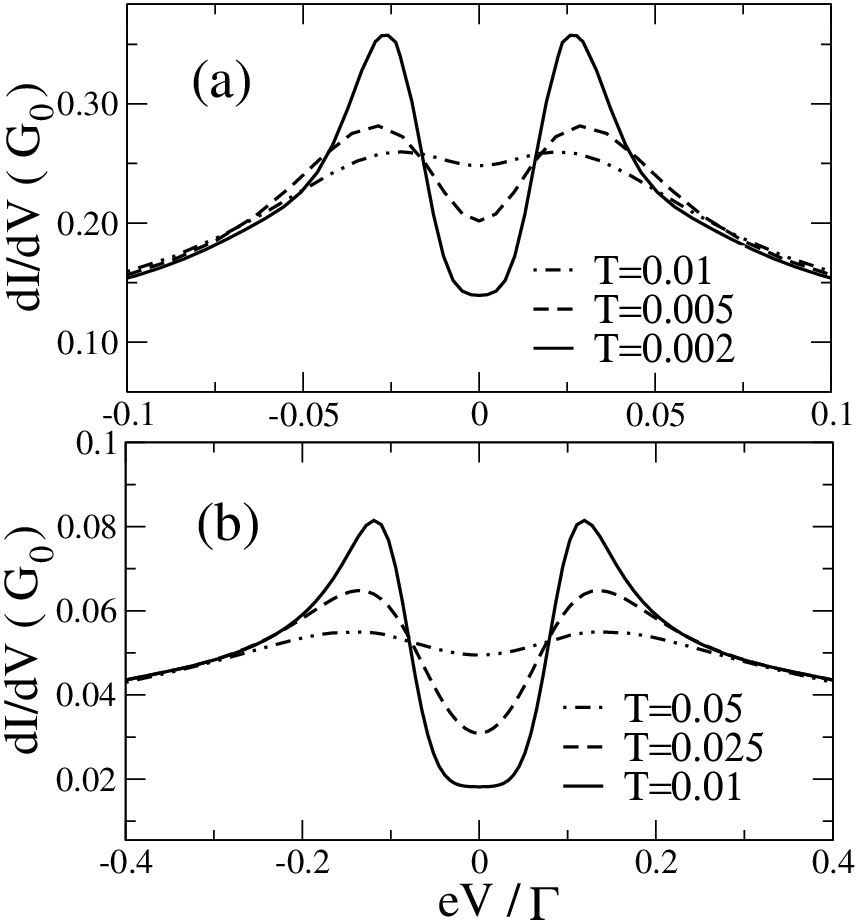}
\caption{Differential conductance as a function of
bias voltage for several temperatures and (a) $E_t=-2, E_s=-2.03$, (b) $E_t=-3, E_s=-3.1$.}
\label{didvs}
\end{figure}

The differential conductance $dI/dV$ as a function of bias voltage on the singlet side
of the transition is
displayed in Fig. \ref{didvs} (a).
We have
applied the voltage symmetrically ($\mu _{L}=-\mu _{R}=eV/2$). As the
temperature is lowered, a dip develops at small voltages. The half width of the
dip is of the order of  $E_{t}-E_{s}$. These results are in
good agreement with the experimental ones (Fig. 4 c of Ref. \cite{roch}). 
For other parameters, in particular larger  values of $E_{t}-E_{s}$ and $E_{d}-E_{s}$
(less valence fluctuations) we obtain curves that look similar to those
reported in carbon nanotubes, with a flat bottom at low temperatures, 
explained using poor man's scaling \cite{paa}.  An example is illustrated in Fig. 
\ref{didvs} (b) 
%for the same $E_{s}$ and $E_{t}$ as in Fig. \ref{gs}.

\begin{figure}[tbp]
%[rbp]
\vspace{1cm}
\includegraphics[width=7cm]{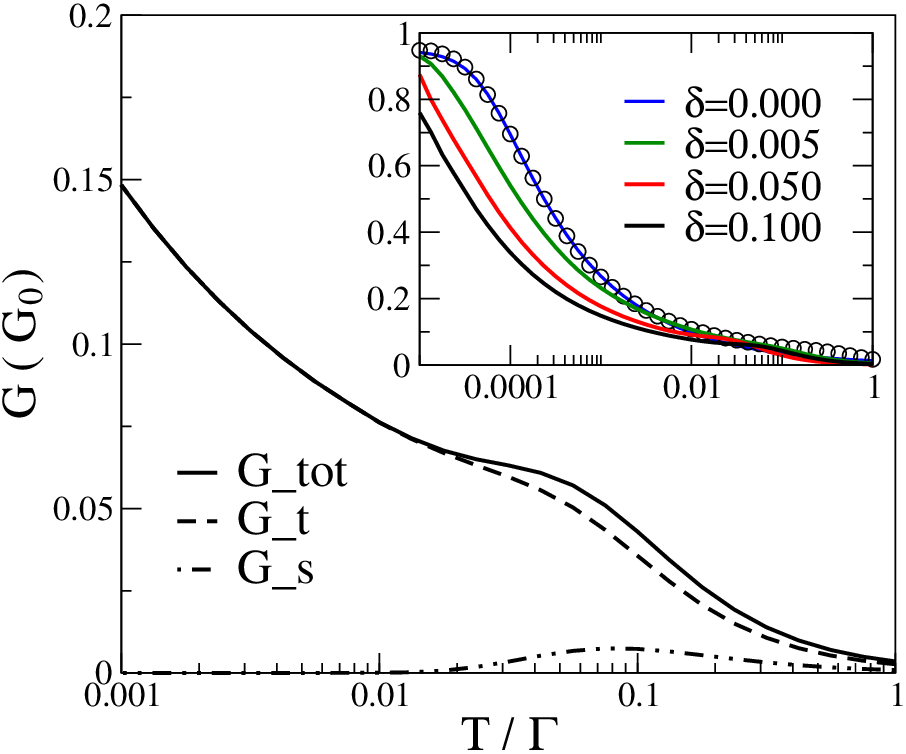}
\caption{(Color online) Zero bias conductance $G(T)$ as a function of
temperature (full line) and contributions from the singlet (dashed dot dot line) and triplet 
(dashed line) for $E_t=-3, E_s=-2.9$. The inset shows $G(T)$ for $E_t=-3$ and 
several values of $\delta= E_s-E_t$ 
%ACA (from top to bottom $\delta=0, 0.005, 0.05$ and $0.1$). 
(increasing from from top to bottom). 
The circles correspond to Eq. (\ref{ge}). $G_{0}=2e^{2}A/h$.}
\label{gt}
\end{figure}

The most novel experimental results are those on the ``triplet side" of the
transition ($E_{s}>E_{t}$). As the temperature is decreased,  $G(T)$
increases until it reaches a plateau at the characteristic energy
$E_{s}-E_{t}$ and then it continues to increase \cite{roch}. This is in
general, the behavior that we obtain, as displayed in Fig. \ref{gt}. As the
partial contribution to the conductance for each density reveals, the
plateau is due to the contribution of the singlet part of the dot spectral density, namely 
$\rho_{d}^{s}(\omega )$  which is peaked at $-(E_{s}-E_{t})$ (see Fig. \ref{dens}).
This plateau has not been observed in previous calculations using NRG \cite{hof}.
We believe that the reason for this is the lack of resolution of NRG to describe
peaks in the spectral density out of the Fermi energy, like that of $\rho_{d}^{s}(\omega )$.
A clear manifestation of this is a system in which the Kondo peak is split in two 
out of the Fermi level \cite{vau}. In fact, while NRG uses a logarithmic
frequency mesh centered at the Fermi energy, we find that in order to obtain 
enough accuracy in the convolutions that define
$\rho _{d}^{s}(\omega )$ and $\rho_{d}^{t}(\omega )$, it is necessary to use 
two different dense logarithmic meshes, centered at the corresponding peaks.

The inset of Fig. \ref{gt} shows the evolution of the equilibrium
conductance as the system is displaced from the quantum critical line to the triplet region.
At this line, $G(T)$ is the same as the corresponding result for the OAM
obtained using the mapping mentioned above. In this case, a fit of $G(T)$
using the empirical curve derived by fitting results of the NRG for a spin 1/2

\begin{equation}
G_{E}(T)=\frac{G(0)}{\left[ 1+(2^{1/s}-1)(T/T_{K})^{2}\right] ^{s}},
\label{ge}
\end{equation}
with $s=0.22$ works very well. We could not find other regions in which
similar empirical curves would fit well a large portion of the curve. When $E_{s}$
increases, the degeneracy in the ground state is reduced and $G(T)$
decreases in the whole range of temperatures. However, in contrast to the
case $E_{s}<E_{t}$, the conductance at zero temperature retains values near
to the ideal one $G_{0}=2e^{2}A/h$ due to the partial Kondo effect.

\begin{figure}[tbp]
%[rbp]
\includegraphics[width=7cm]{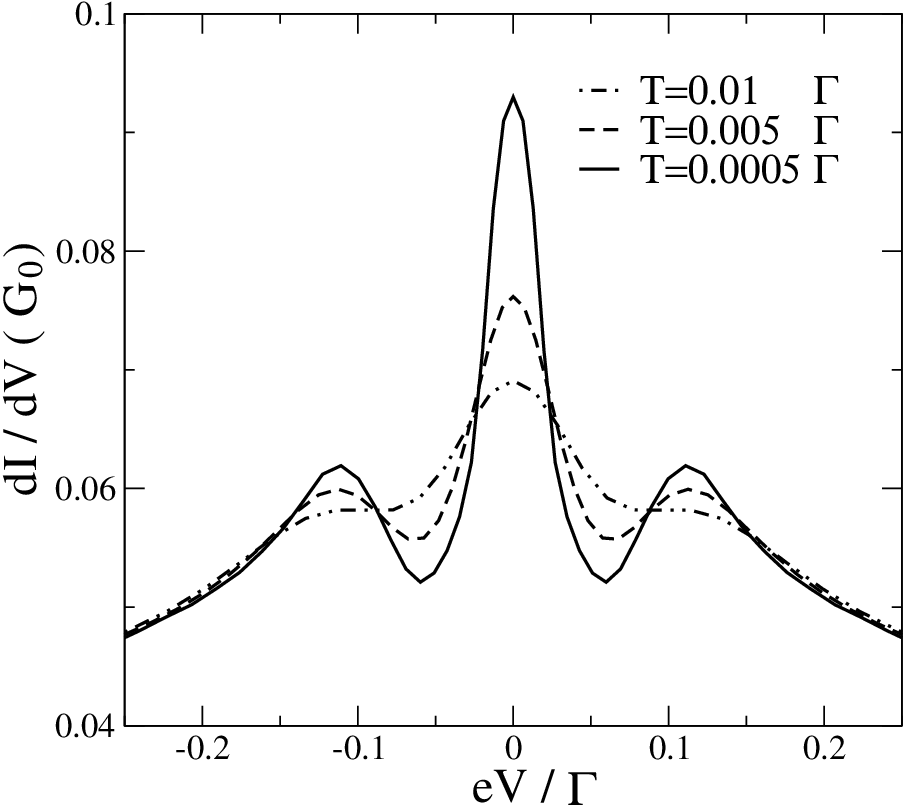}
\caption{Differential conductance as a function of
bias voltage for $E_t=-3, E_s=-2.9$ and several temperatures.}
\label{didvt}
\end{figure}

In Fig. \ref{didvt} we show the non-equilibrium differential conductance 
$dI/dV$  on the triplet side at several temperatures. At low temperatures, 
there is one peak centered at $V=0$ and other two centered at $V= \pm V_M$. 
This three-peak structure also agrees qualitatively well with experiment \cite{roch}. 
Actually in the later, the peak at $V= - V_M$ is higher than that at $V= V_M$, while
our results are even in $V$ due to our assumption of a symmetric 
voltage drop. Changing this we can easily control the relative height
of both peaks. In any case, the main point is the existence of these three peaks,
which is also a consequence of the particular
structure of the spectral densities  $\rho _{d}^{s}(\omega )$ and
$\rho_{d}^{t}(\omega )$ at equilibrium and low
temperatures (see Fig. \ref{dens}). The peak at $V=0$ is due to the
$\rho_{d}^{t}(\omega )$ which is peaked at the Fermi energy.
%and broadens and loses intensity as the voltage $|V|$ is increased. 
When $e|V|$ reaches energies at which $\rho _{d}^{s}(\omega )$ is peaked, 
this density starts to contribute significantly to the current $I$ [see Eq. (\ref{i})] and
$dI/dV$ increases. Finally, for large $e|V|$, $I$ tends to saturate and 
$dI/dV$ decreases again.

The effect of temperature is to broaden the peaks and at high enough temperatures only 
one broad peak in $dI/dV$ is present, in agreement with experiment.
For different parameters as those of Fig. 1, this broadening of the spectral densities 
is already important at
$V=T=0$ and increases with $|V|$ in such a way that even at $T=0$ only one peak
in $dI/dV$ is present. This
happens for example at  $E_{t}=-2\Gamma $, $E_{s}-E_{t}=0.03\Gamma $, for
which $dI/dV$  (not shown) has a monotonic behavior for positive $V$,
displaying only a shoulder for $eV\sim \pm (E_{s}-E_{t})$. 

In summary, the transport properties recently observed near the singlet-triplet
quantum phase transition in quantum dots can be explained in the framework of the singlet-triplet 
Anderson model with one channel per conduction lead, out of equilibrium, using the non-crossing
approximation. We made use of exact results to control the distance to the
quantum critical point. The differential conductance $dI/dV$ as a function
of bias voltage is markedly different at both sides of the transition
showing a dip (peak) at small voltages on the singlet (triplet) side and
often a three-peak structure on the triplet side. The zero bias
conductance $G(T)$ as a function of temperature displays a plateau due to
the contribution of the excitations from the localized singlet to the
doublet. 
%$\rho _{d}^{s}(\omega )$. 

\section*{Acknowledgments}

We thank Ana M. Llois for useful discussions. 
One of us (AAA) is supported by CONICET. This work was done in the framework
of projects PIP 5254 and PIP 6016 of CONICET, and PICT 2006/483 and 33304 
of the ANPCyT.

\end{document}